\documentclass{eptcs}
\providecommand{\event}{VORTEX 2019} 
\usepackage{breakurl}             

\usepackage[utf8]{inputenc}
\usepackage[T1]{fontenc}
\usepackage[english]{babel}
\usepackage{microtype} 
\usepackage{tabularx} 
\usepackage{booktabs} 
\usepackage{lmodern} 

\usepackage[mode=buildnew]{standalone}
\usepackage{amsmath} 
\usepackage{amssymb}
\usepackage{marginnote}
\usepackage{lmodern}
\usepackage{textcomp}
\usepackage{courier}
\usepackage{graphicx}
\usepackage{subcaption}
\usepackage{tikz}
  \usetikzlibrary{snakes,arrows,shapes,matrix, shapes.misc, decorations, decorations.pathreplacing, automata}
\usepackage{tikz-timing}
\tikzset{
  timing caption/.style={
    font=\scriptsize,
    minimum height=3ex,
    inner sep=0pt
  }
}

\usepackage{paralist}
\usepackage{xspace}
\usepackage[para]{footmisc}
\usepackage[english=american]{csquotes}
\usepackage{hyperref}
  \hypersetup{breaklinks=true,
              pdfborder={0 0 0},
              pdfhighlight={/N}}
\usepackage[all]{hypcap}
\usepackage{todonotes}
\usepackage{listings}
\usepackage{booktabs}
\usepackage[capitalise]{cleveref}
\usepackage{acronym}

\usepackage{wrapfig}
\usepackage{stmaryrd}

\usepackage[characters]{ltl}
\usepackage{import}

\newcommand{\op}[1]{\operatorname{#1}}
\newcommand{\www}[1]{\href{http://#1}{\nolinkurl{#1}}}

\newcommand{\LTLprophecy}[1]{\triangleright_{[#1]}}

\newcommand{\LTLprophecyobligation}[1]{\triangleright'_{[#1]}}

\let\meet\sqcap
\let\join\sqcup

\let\phi\varphi

\providecommand{\parag}[1]{\par\noindent\emph{#1}.\ }
\providecommand{\sect}[1]{\par\noindent\textbf{#1}.\ }

\newcommand{\circled}[1]{
  \setbox0=\hbox{#1}%
  \dimen0\wd0%
  \divide\dimen0 by 2%
  \tikz[baseline=(a.base)]{%
      \useasboundingbox (-\the\dimen0,0pt) rectangle (\the\dimen0,1pt);%
      \node[circle,draw,outer sep=0pt,inner sep=0.1ex] (a) {#1};%
  }\,}

\newcommand{\circledblue}[1]{
  \setbox0=\hbox{#1}%
  \dimen0\wd0%
  \divide\dimen0 by 2%
  \tikz[baseline=(a.base)]{%
      \useasboundingbox (-\the\dimen0,0pt) rectangle (\the\dimen0,1pt);%
      \node[circle,blue,draw,outer sep=0pt,inner sep=0.1ex] (a) {#1};%
  }\,}

\newcommand{\bluecircled}[1]{
  \setbox0=\hbox{#1}%
  \dimen0\wd0%
  \divide\dimen0 by 2%
  \tikz[baseline=(a.base)]{%
      \useasboundingbox (-\the\dimen0,0pt) rectangle (\the\dimen0,1pt);%
      \node[circle,draw,outer sep=0pt,inner sep=0.1ex,white,fill=blue] (a) {#1};%
  }\,}

\newcommand{\evaluateMark}{\raisebox{.2em}{\tiny\color{blue}\checkmark}}
\usetikzlibrary{arrows,backgrounds,positioning,calc,intersections,shapes}

\tikzset{box/.style={inner sep=1pt, minimum height=1.2em}}
\tikzset{node/.style={box, minimum width=9ex, semithick, rectangle}}
\tikzset{split/.style={rectangle split, rectangle split parts=2, rectangle split draw splits=true}}
\tikzset{split2/.style={rectangle split, rectangle split horizontal, rectangle split parts=2, rectangle split draw splits=true}}
\tikzset{arrow/.style={->, >=latex', semithick}}
\tikzset{label/.style={auto,midway,above}}

\tikzset{system/.style={}}
\tikzset{system-split/.style={system, split}}
\tikzset{agent/.style={thick}}
\tikzset{task/.style={node, fill=black!10, draw}}
\tikzset{initial/.style={inner sep=0pt, minimum size=3mm, circle, fill=black, draw}}
\tikzset{input/.style={node, fill=red!10, draw=red, text=red}}
\tikzset{message/.style={fill=blue!10, draw=blue, text=blue}}
\tikzset{messageguard/.style={box, blue}}
\tikzset{timedguard/.style={box, draw=blue, text=blue}}
\tikzset{timedguard-split/.style={timedguard, split, draw=blue, text=blue}}

\tikzset{transition/.style={arrow}}
\tikzset{message-1/.style={semithick, dashed, blue}}
\tikzset{message-2/.style={arrow, dashed, blue}}
\tikzset{input-1/.style={semithick, dashed, red}}
\tikzset{timed-1/.style={semithick, dashed, blue}}

\tikzset{label/.style={auto,midway,above}}
\tikzset{label-input/.style={label,text=red}}
\tikzset{label-message/.style={label,text=blue}}

\title{Empowering Multilevel DSMLs with Integrated Runtime Verification}
\def\titlerunning{Empowering Multilevel DSMLs with Integrated Runtime Verification}
\author{Fernando Macías, Adrian Rutle, Volker Stolz\institute{Western Norway University of Applied Sciences, Bergen, Norway\thanks{This work was
    partially supported by the European Horizon 2020 project COEMS under grant agreement no.~732016 (\url{https://www.coems.eu/}).}}
\and Torben Scheffel, Malte Schmitz\institute{Institute for Software Engineering and Programming Languages, University of Lübeck, Germany}}
\def\authorrunning{Fernando Macías \textit{et al.}}

\begin{document}

\maketitle

\begin{abstract}
Within Model-Driven Software Engineering, Domain-Specific Modelling has proven to be a powerful technique to specify systems and systems' behaviour in a formal, yet understandable way.
Runtime verification (RV) has been successfully used to verify the correctness of such behaviour. 
Specifying behaviour requires managing various levels of abstractions,
making multilevel modelling (MLM) a suitable approach for this task.
In this paper, we present an approach to combine MLM and RV with an example from the domain of distributed real-time systems.
The semantics of the specified behaviour as well as the evaluation of correctness properties are given by model transformation rules. 
This facilitates simulation of the system and checking against real-time temporal logic correctness properties.
\end{abstract}

\section{Introduction and Background}
\label{sec:introduction}

Model-driven software engineering (MDSE) is one of the emergent responses from the scientific and industrial communities to tackle the increasing complexity of software systems.
MDSE utilises abstractions for modelling different aspects of software systems, and treats models as first-class entities in all phases of software development.
There are quite a few studies which support gains in quality, performance, effectiveness, etc. as a result of using MDSE (see, e.g.,~\cite{WhittleHRBH13,WhittleHR14,BurdenHW14,MohagheghiGSFNF13,MohagheghiGSF13}).
However, using modelling to understand a domain, making the right abstractions, and including all the stakeholders in the development process, are without doubt the main gains of MDSE~\cite{BurdenHW14,MohagheghiFMFG08,Tolvanen016}.

According to empiric evaluations related to the status and practice of MDSE, the state-of-the-art modelling techniques and tools do a poor job in supporting software development activities~\cite{WhittleHR14}.
One way to increase the adoption of MDSE in practice is to develop modelling approaches which reflect the way software architects, developers and designers, as well as organisations, domain experts and stakeholders, handle abstraction and problem-solving.
We believe that domain-specific (meta)modelling (DSMM)~\cite{deLara2015,MohagheghiH10,Tolvanen016,Kelly08DSMBook} is an approach that could unite software modelling and abstraction, software design and architecture, and organisational studies~\cite{WhittleHR14}.

DSMM is the art of creating and using languages which are specifically tailored for a particular domain~\cite{deLara2015,fowler2011dsl}.
In this case, the concept of ``domain'' corresponds to both real-life (problem) domains --- such as banking, robotics, healthcare, product line systems, education, etc. --- and technical (solution) domains --- such as SQL, HTML, SysML, Petri Nets, etc.
The idea of DSMM is already practiced in industry and academia and its positive results are documented in the literature~\cite{WhittleHR14,BurdenHW14,MohagheghiGSFNF13}.
Moreover, mainstream approaches to DSMM and design and implementation of domain-specific languages (DSL), such as the Eclipse Modelling Framework~\cite{eclipse_modeling_framework_web}, the Unified Modelling Language, MetaCase~\cite{metacase}, etc., are all based on two-level (meta)modelling approaches.
That is, domain concepts are defined in metamodels, and they are instantiated in models which conform to these metamodels.
This also includes the usage of so-called profiles' or stereotypes'.
Provided that in the real-life domains the way of thinking is not limited to a certain number of abstraction levels, approaches which force designers to adapt their way of thinking are deemed to fail.
Furthermore, describing software domains only in terms of models and their instances usually leads to unnecessary complexity and synthetic type-instance relations~\cite{atkinson2008reducing}.
In this work, we propose an alternative approach based on multilevel modelling (MLM) for DSMM, so that the number of abstraction levels are not limited.

In addition to structure, modelling systems' behaviour is necessary in order to fully understand how they work in practice.
Modelling behaviour is inherently multilevel since the behavioural modelling language is defined by a metamodel while the semantics is described two levels below the metamodel~\cite{deLara2015,delara2010mixin,Rutle:2012:MAB}.
The cause is that behaviour is reflected in the running instances of the models which in turn conform to their metamodel.
Hence, MLM hierarchies could be used for DSMM, i.e., for the flexible organization of metamodelling languages, models and their running instances~\cite{deLara2015}.

Correctness of behaviour can be addressed through model checking, which usually requires particular tools, and addressing issues such as state-space explosion.
As we would like to keep expressive modelling and some degree of verification within the same framework so that we can link the system and its properties,
we employ Runtime Verification (RV) as a lightweight formal method, which checks one execution of the system against a temporal correctness property.
By representing both system and properties as models belonging to \textit{related} MLM hierarchies, we can simulate different executions of the system and check them against the specified properties, while treating these two aspects of the system --- behaviour and correctness --- uniformly.
 
To explain our approach, we use distributed real-time systems (in particular, a robotic system) as a sample modelling domain --- which is complex enough yet easy to grasp --- together with linear temporal logic to define correctness properties.
We will keep the behavioural models and the correctness specification synchronized by connecting the propositions used in the correctness properties to the models in a MLM hierarchy.
The semantics of both the modelled behaviour and the evaluation of the correctness properties are given by the application of model transformation rules%
These rules are executed on the models and the properties simultaneously --- creating and verifying execution paths in the state space of the system --- using a uniform technique.

We start with an explanation of the logic which we use for property specification in the domain of distributed real-time systems in Section~\ref{sec:ltl}.
Section~\ref{sec:combine-model-and-properties} describes how we combine MLM and RV in our setting, Section~\ref{sec:mcmts} contains how the simulation with MCMTs works and Section~\ref{sec:ftltl-modelling-semantics} explains how we can evaluate RV properties as models using model transformation rules.
Related work, conclusions and future lines of research are presented in Section~\ref{sec:related}.

\section{Timed Linear Temporal Logic}
\label{sec:ltl}

Correctness properties for RV are usually specified in a temporal logic like the Linear Temporal Logic (LTL).
With LTL one can specify properties about the future of a single execution trace of the system under scrutiny.
Such an execution trace is represented as a discrete sequence of execution states, i.e. a finite word $w \in \Sigma^+$.
In every state, every \textit{atomic proposition} of the finite set $\op{AP}$ is either true or false, i.e. $\Sigma = 2^{\op{AP}}$.
With LTL one can specify occurrence and ordering constraints, but LTL can be extended to timed LTL (TLTL, \cite{tltl}) to express real-time timing constraints.
Now words are still discrete sequences, but every symbol is paired with a timestamp, i.e. $w \in (\Sigma \times \mathbb T)^+$ for a time domain $\mathbb T$.
Special extensions of LTL are used to apply LTL to asynchronous distributed systems (or agents), e.g. with the Distributed Temporal Logic (DTL, \cite{dtl}) one can specify on which agent subformulas are evaluated.

In this paper we consider asynchronous distributed systems without a shared global clock.
We combine a four-valued semantics for TLTL with a simplified version of DTL, where remote subformulas are restricted to atomic propositions, to be able to specify and monitor real-time constraints on the different agents.
TLTL can express properties about relative time differences through the \emph{prophecy} operator $\triangleright$.
We restrict TLTL to the future fragment which can only relate current events with upcoming events.
This fragment can be monitored using term-rewriting techniques known as unrolling \cite{DBLP:conf/kbse/HavelundR01,ltl4}.
Now let $\op{AP}$ be the set of atomic propositions which are evaluated either to $\LTLtrue$ or $\LTLfalse$ in every event. Then $w = (a_0,t_0),\dots,(a_n,t_n) \in (\Sigma \times \mathbb{T})^+$ is a finite event sequence over the alphabet $\Sigma = 2^{\op{AP}}$ with time domain $\mathbb T$. For such a word $w$ we denote with $w^i := (a_i,t_i)\ldots(a_n,t_n)$ the suffix starting with event $a_i$.
The syntax of TLTL formulas $\varphi$ is given by the following grammar for $p \in \op{AP}$ and $t_1,t_2\in \mathbb{T}$ with $t_1 < t_2$:
$\varphi ::= p \mid \neg \varphi \mid \varphi \lor \varphi \mid \LTLnext \varphi \mid \varphi\LTLuntil\varphi\mid \LTLprophecy{t_1,t_2} p$.
For formulas $\phi$ and $\psi$ and a word $w$, the semantics for a TLTL formula is given inductively as a relation $\models$ between words and the formulas they fulfil:

\begin{minipage}[t]{0.3\linewidth}
$\begin{array}{l}
w \models p \Leftrightarrow  p \in a_0 \\
w \models \neg \varphi \Leftrightarrow w \not\models \varphi\\
w \models \varphi \lor \psi \Leftrightarrow w \models \varphi \text{ or } w \models \psi \\
w \models \LTLnext\varphi \Leftrightarrow |w| > 1 \land w^1 \models \varphi\\
\end{array}$
\end{minipage}
~~~~~~~~~~~\begin{minipage}[t]{0.45\linewidth}
$\begin{array}{l}
w \models \varphi \LTLuntil \psi \Leftrightarrow  \exists i: w^i \models \psi \land \forall j < i: w^j \models \varphi\\
w \models \LTLprophecy{t_l,t_h} p \Leftrightarrow \exists i: p \in a_i \land t_l \leq t_i \leq t_h\\
 ~~~~~~~~~~~~~~~~~~~~ \land \forall j: 0 < j < i \to p \notin a_j
\end{array}$
\end{minipage}

\noindent Additional LTL operators can be added as syntactic sugar, e.g. through the following equivalences:
$\LTLtrue :\equiv \varphi\lor\neg\varphi$,\quad
$\LTLfalse :\equiv \neg \LTLtrue$,\quad
$\varphi \land \psi :\equiv \neg(\neg\varphi\lor\neg\psi)$,\quad
$\LTLfinally\varphi :\equiv \LTLtrue\LTLuntil \varphi$,\quad
$\LTLglobally\varphi :\equiv\neg\LTLfinally\neg\varphi$,\quad
$\LTLweaknext \varphi \equiv \neg \LTLnext\neg\varphi$.
The syntax and semantics can easily be extended to allow $\LTLprophecy{t_l,t_h} \LTLnot p$ through exchanging $p \in a_i$ and $p \notin a_i$.

\sect{Monitoring TLTL}
To evaluate a word against a TLTL formula a step-wise monitoring function $e$ is used. $e$ takes an input symbol (the set of atomic propositions valid in that event), the relative time elapsed since the last event and the current formula.
$e$ returns an intermediate verdict indicating if the formula is fulfilled by the word up to this event and the rest of the formula which has not been fulfilled yet.
We use the truth domain $\mathbb B_4 = \{\top, \top^c, \bot^c, \bot\}$ for the output verdicts, where $\top$ and $\bot$ represent final verdicts and $\top^c$ and $\bot^c$ represent current verdicts up to now, which might change with new events arriving \cite{ltl4}.
Because $\mathbb B_4$ is a lattice it provides the operators \emph{meet} $\meet$, \emph{join} $\join$ and \emph{complement} as generalization of the boolean operators \emph{and} $\wedge$, \emph{or} $\vee$ and \emph{not}, e.g. $\top \meet \bot^c = \bot^c$, $\top \join \bot^c = \top$ and $\overline{\bot^c} = \top^c$.

\sect{Monitoring Distributed Systems}
We adopt the concepts behind DTL as presented in \cite{dtl} to get a simplified monitoring of the behaviour of a distributed system: 
The formula $@_A\phi$ represents the formula $\phi$ being evaluated on agent $A$. 
The TLTL formula $\phi$ can contain other $@$-annotations followed by atomic propositions, which refers to the current evaluation of that proposition on that agent. Note that as in DTL we only specify properties with respect to the knowledge that an agent has about the other agents. 

\section{Combining Models and Properties}
\label{sec:combine-model-and-properties}

In MDSE, models are commonly formalised using concepts from graph and category theory, which is also applicable to multilevel modelling.
In this paper, we use a graph-based formalisation that defines multilevel hierarchies of models as graphs whose elements have typing relations among them.
That is, every element in a model has a type which must be defined in a model at a higher level of abstraction.
Models belong to levels, which are numbered incrementally from level 0, which contains the self-defining root model of the hierarchy.

\subsection{Robolang hierarchy}

\begin{figure}
	\centering
	\includegraphics[width=.8\textwidth]{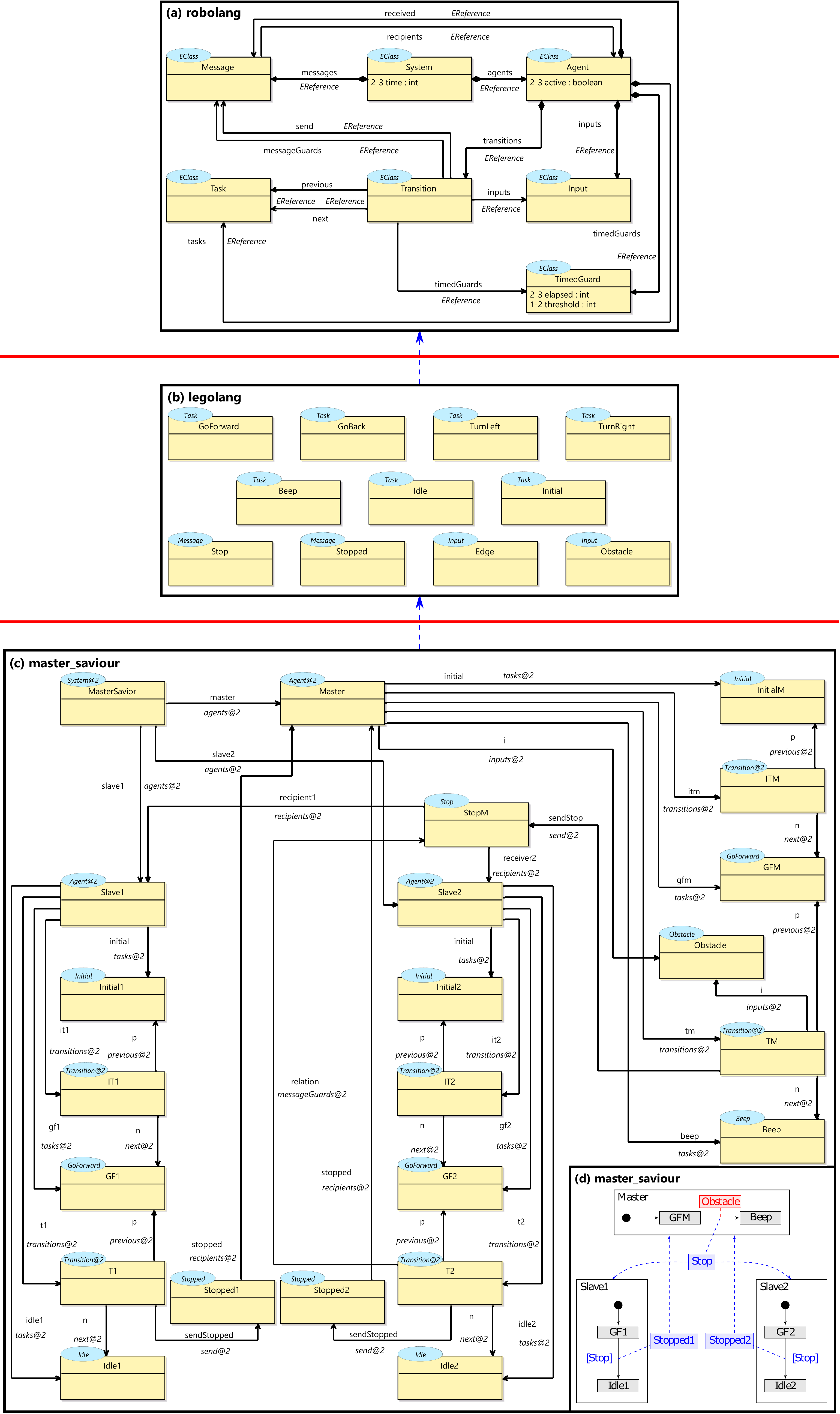}
	\caption{Robolang MLM hierarchy for definition of robot behaviour}
	\label{fig:hierarchy}
\end{figure}

The MLM hierarchy used in this paper depicts a DSML which enables modelling of distributed real-time robots (see Fig.~\ref{fig:hierarchy}). 
It allows to specify different agents, each with its own independent workflow, as well as the messages that they send each other when firing a transition.
These messages can also act as guards of a transition, so that it is only fired once the message is received.
Additionally, the concepts of explicit time passing and timeout are part of the models.
In this hierarchy, elements are visualised as rectangles and their types are depicted with a blue ellipse on the rectangle, while relations are visualised as arrows with their types in italics font style.

The top-most model of Fig.~\ref{fig:hierarchy}(a), \textsf{robolang}, contains the elements \textsf{Task}, \textsf{Transition} and \textsf{Input}, and the references among them; as seen from the types \textsf{EClass} and \textsf{EReference}, this model is typed over Ecore~\cite{steinberg2008emf}. 
These three elements are contained inside \textsf{Agent}, a concept used to represent each of the robots which participate in a scenario.
Agents have an \textsf{active} boolean attribute which indicates if the robot has switched to a new state in the last time step or remained in its current one, e.g. by firing a transition or receiving a message.
Agents can also contain \textsf{TimedGuards}.
The two attributes inside a timed guard are meant to be instantiated in adjacent levels: when designing a specific workflow, the \textsf{threshold} attribute must be given a value which indicates the minimum amount of time that the task prior to a transition must be executed before the transition (which connects to the guard via the \textsf{timedGuards} arrow) can be fired.
In the levels below, the \textsf{elapsed} attribute is used to keep track of the amount of time that the task has been running already.
Once this number become greater or equal than the threshold specified in the level above, the transition can be fired.
The elapsed time in timed guards must increase in the same amount as the global \textsf{time}  in \textsf{System}, which represents the whole modelled scenario and contains all agents.

As shown in the figure, \textsf{Message} can be contained by either the system (if the message was just sent) or by an agent (if the message was received already).
A transition can send the messages indicated by the \textsf{send} relation when it is fired to the agents that the message connects to, using the \textsf{recipients} relation.
When the message is received, it can trigger other transitions to which it is connected using the \textsf{messageGuard} relation.

In our case, we are working with simple Lego robots hence we adapt the language \textsf{robolang} to cope for specific concepts and behaviour.
The model \textsf{legolang} in Fig.~\ref{fig:hierarchy}(a) contains the refinements of task, input and message into more specific ones. 
These refinements are performed by creating a new modelling level and adding it to the hierarchy below \textsf{robolang}.
Elements of the new model are typed by elements of the original model, as indicated by the blue ellipses.
We specify here four types of movement in two dimensions for our robots: \textsf{GoForward}, \textsf{GoBack}, \textsf{TurnLeft} and \textsf{TurnRight}.
There are also tasks to make the robot \textsf{Beep}, stay \textsf{Idle} and the special task \textsf{Inital} which marks the starting point of the workflow.
There are also the inputs \textsf{Edge} and \textsf{Obstacle}, and two types of message: \textsf{Stop}, where the sender orders the receiver to remain idle, and \textsf{Stopped}, as an acknowledgement for it.
Other refinements of the \textsf{robolang} model will give different branches with a common root, hence yielding a tree-shaped hierarchy.


The \textsf{master\_saviour} model in Fig.~\ref{fig:hierarchy}(a) shows a relatively simple scenario which uses most of the concepts in this version of the language.
Although it is machine readable, the syntax shown in this model is not human-friendly.
Therefore, we propose a concrete syntax for describing scenarios in Fig.~\ref{fig:hierarchy}(d) using the following representations for the instances of different types:

\begin{itemize}
	\item The instance of system containing the other elements has no explicit representation, but simply becomes the canvas in which the rest of the graphical elements are laid out.
	\item Instances of agents appear as black boxes.
	\item Instances of tasks are represented as grey rectangles except the initial ones which are represented with a black dot.
	\item Instances of transitions appear as arrows between previous and next tasks.
	\item Instances of inputs are red boxes, connected by a dashed line to the transition they trigger.
	\item Instances of messages which are sent appear as blue boxes, connected to the sender transitions and to their receivers by dashed blue lines. Messages that act as guards appear as a blue label for the transition they trigger and within square brackets.
\end{itemize}

Keeping this representation in mind, it is now easier to understand the modelled scenario:
There are three agents, which we can imagine moving in the same direction.
One of them, the master, has a sensor able to detect obstacles, whereas the other two agents, the slaves, are blind and moving slightly behind their master.
In case there is an obstacle in the way, the master must alert the slaves so that they can stop before crashing into it, and then beep as a notification that an obstacle was found.
The slaves must just keep going forward until they receive the \textsf{Stop} message, to which they respond by sending an instance of the \textsf{Stopped} message each, and then become idle.
Note that the types of the tasks and messages are not visible in this representation, but can be checked in the abstract one, e.g. \textsf{GF1} is an instance of \textsf{GoForward}.

\subsection{Supplementary Hierarchies and Double-Typing}
\begin{figure}
	\centering
	\includegraphics[width=0.8\textwidth]{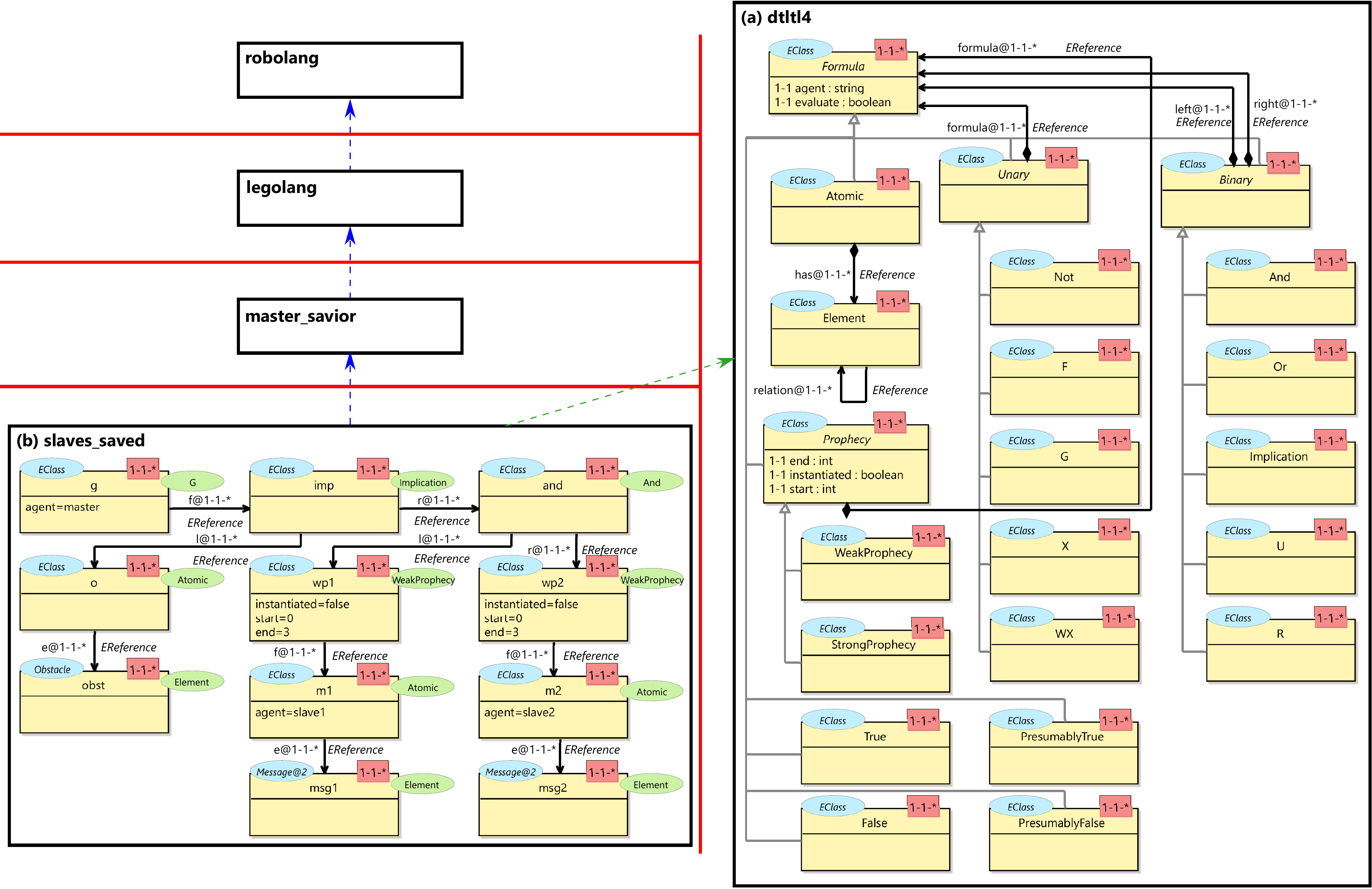}
	\caption{TLTL property as model and TLTL language as supplementary hierarchy}
	\label{fig:dtltl-multilevel-hierarchy}
\end{figure}

The hierarchy \textsf{Robolang} contains the language and the distributed scenario with the robots that we want to simulate.
We call this for \textit{application hierarchy}, depicting that the application domain is definition of robotic models.
This hierarchy is independent from the \textsf{TLTL} hierarchy that contains the syntax for the TLTL language.
The \textsf{TLTL} hierarchy is shown in Fig.~\ref{fig:dtltl-multilevel-hierarchy} and explained shortly in Section \ref{sec:ftltl-modelling-semantics}.
In order to relate both hierarchies and be able to specify temporal properties for the simulated scenario, we exploit the concept of \textit{supplementary hierarchy}~\cite{macias2016emisa}.
By declaring one hierarchy as supplementary to the application hierarchy, elements in the latter can be double-typed with supplementary types.
The supplementary typing relations are indicated in green to differentiate them from the ones in the application hierarchy (in blue).
Now we can define temporal properties in which the atomic propositions are small model fragments, which are at the same time part of the TLTL property (due to their supplementary type) and consistent with the types of \textsf{Robolang} (due to their application type).
The model in Fig.~\ref{fig:dtltl-multilevel-hierarchy}(b) represents the property ``when master detects an obstacle, both slaves will get messages within 3 time units'', explained in Section~\ref{sec:ftltl-modelling-semantics}.

\section{Simulation Semantics with MCMTs}
\label{sec:mcmts}

The semantics for models that represent behaviour, like the ones in the example, can be implemented using model transformations.
We will use Multilevel Coupled Model Transformations (MCMT, \cite{macias2016emisa}) rules for both the simulation of the system and the evaluation of TLTL properties which the system should satisfy.
MCMTs are a kind of model transformations which use multilevel information to increase the flexibility and reusability of model transformations.
MCMT rules are expressed in a declarative manner, with a \textit{FROM} block that needs to match (i.e. be isomorphic and compatible with multilevel typing) against the model in order to be applied.
If such a match found, the matched subgraph is replaced by the one specified in another block, called \textit{TO}.
The information about multilevel typing is specified in a different block called \textit{META}, which requires to be matched in the metalevels of the model being transformed, prior to the match of \textit{FROM}.

In the \textsf{Robolang} case, MCMTs describe the simulation semantics used to generate the sequence of models which simulate the evolution of the system through time.
While in the \textsf{TLTL} case, MCMT rules are used to monitor the simulation, performing unrolling and evaluation on the model which represents the TLTL property. These rules implement the TLTL monitor function $e$ from Section~\ref{sec:ltl}.
Both sets of MCMT rules are then coordinated to be run with specific priorities, so that the simulation and the evaluation can be performed in an interleaving fashion, as illustrated in the examples in Fig.~\ref{fig:example-execution} and~\ref{fig:example-evaluation} and described in more detail in the next section.

Using MCMTs we can specify the complete semantics of the \textsf{Robolang} language using a coordination algorithm and four groups of rules: \textit{Behavioural}, \textit{Environmental}, \textit{Time stepping} and \textit{Remove active marks}.
The coordination algorithm is given at the end of this section and defines the number of rules and the order in which the rules from each group are executed.
In the following, we explain the most relevant MCMTs, using the concrete syntax of our example in Fig.~\ref{fig:hierarchy}, and an explanation of their semantics.
For simplicity, instead of the blue ellipses we use \textsf{E:T} to indicate that an element \textsf{E} has a type \textsf{T}.

\subsection{Behavioural Rules}
\label{subsec:behavioural-rules}

The \textit{behavioural} rules encode the semantics of how agents evolve through time, execute tasks, switch from one task to the next, and interact with messages and environmental information.

\begin{wrapfigure}[9]{r}{0.5\textwidth}
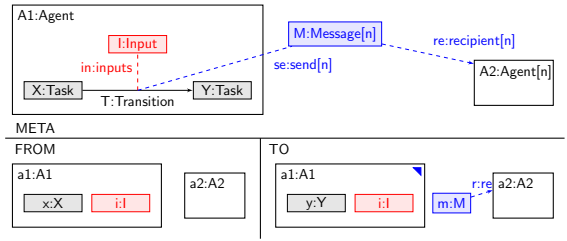

    \vspace{-5mm}
    \centering
	\scalebox{.5}{\includestandalone{images/robolang-mcmt-fire-transition-with-input}}
	\caption{MCMT Fire Transition with Input}
	\label{fig:robolang-mcmt-fire-transition-with-input}
\end{wrapfigure}
\parag{Fire Transition with Input} (Fig.~\ref{fig:robolang-mcmt-fire-transition-with-input}) 
This rule allows an agent to progress by responding to environmental inputs (red).
If a message has to be sent during the transition, messages into the system (outside any agent) are created.
One message per receiving agent is created, with exactly one reference to the sending agent and exactly one reference to the recipient (blue).
Note that due to the \textit{[n]} multiplicity specifier this rule will match scenarios in which \textit{A1} sends any number of messages to the same number of agents.

\begin{wrapfigure}[9]{r}{0.5\textwidth}
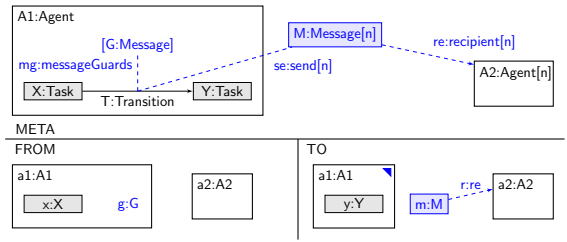

    \vspace{-5mm}
    \centering
	\scalebox{.5}{\includestandalone{images/robolang-mcmt-fire-transition-with-message-guard}}
	\caption{MCMT Fire Transition with Guard}
	\label{fig:robolang-mcmt-fire-transition-with-message-guard}
\end{wrapfigure}
\parag{Fire Transition with Message Guard} (Fig.~\ref{fig:robolang-mcmt-fire-transition-with-message-guard})
This rule is similar to the previous one, but the transition is triggered by the reception of a message, which acts as a guard.
As opposed to inputs, guards are consumed when the transition is fired.
That means that this MCMT removes the guard from the running instance (in the TO block) once it is fired.

Two similar rules for firing both the initial transitions and those triggered by a timed guard, together with a simple rule for adding them, have been excluded due to space limitations and their absence in the example.

\subsection{Environmental Rules}
The rules from this group are applied to simulate changes on the environment of the agents, such as a change perceived through some input from the agent's sensors or a message being received from another agent.

\begin{figure}
	\centering
        \begin{minipage}{0.35\linewidth}
          \centering
          \includestandalone[scale=0.7]{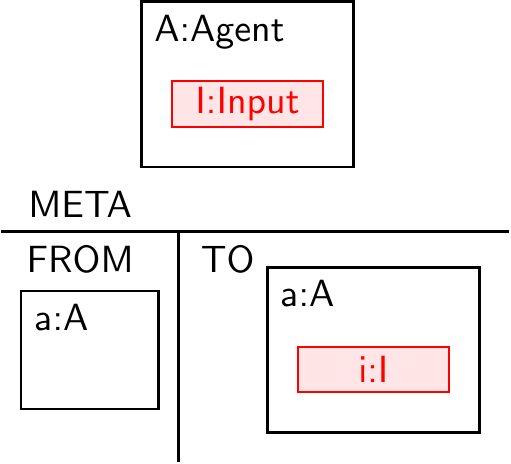}
          \vspace*{8ex} 
          \caption{MCMT Insert Input}
          \label{fig:robolang-mcmt-insert-input}
        \end{minipage}
        \begin{minipage}{0.5\linewidth}
          \centering
          \includestandalone[scale=0.7]{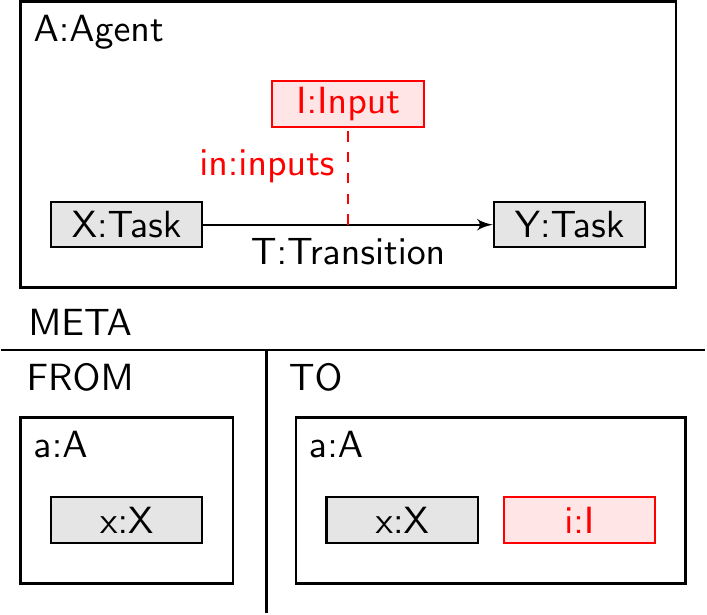}
          \caption{MCMT Insert Effective Input}
          \label{fig:robolang-mcmt-insert-effective-input}
        \end{minipage}
    	\vspace{-4mm}
\end{figure}

\parag{Insert Input} (Fig.~\ref{fig:robolang-mcmt-insert-input}) 
This rule creates an input inside an agent. 
It is a non-deterministic and loose version of the next rule, since the appearance of such input may not cause any effect on the execution.

\parag{Insert Effective Input} (Fig.~\ref{fig:robolang-mcmt-insert-effective-input})
This rule creates an input inside an agent, ensuring that it will cause its currently running task to stop, since it will introduce and input that will allow the rule \textit{Fire Transition with Input} to be executed next.
Although this rule is more restrictive than \textit{Insert Input}, it is still non-deterministic, since there may be more than one agent in which this rule can be matched.

\parag{Delete Input}
Represents the deletion of a previously existing input.
This rule reflects the fact that an input, previously detected by an agent, can not be detected any more.
For instance, if an agent is moving backwards, an obstacle detected in front of it should eventually disappear.
This rule is a mirrored version of \textit{Insert Input} in Fig.~\ref{fig:robolang-mcmt-insert-input}.

\begin{wrapfigure}[9]{r}{0.4\textwidth}
    \vspace{-5mm}
    \centering
	\includestandalone[scale=0.7]{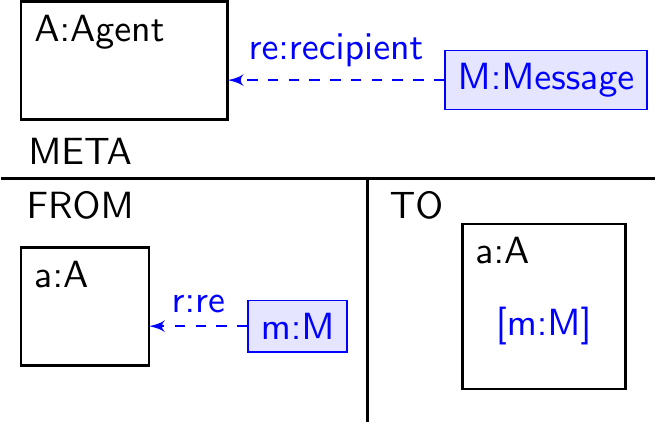}
    \caption{MCMT Receive Message}
    \label{fig:robolang-mcmt-receive-message}
\end{wrapfigure}
\parag{Receive Message}  (Fig.~\ref{fig:robolang-mcmt-receive-message})
If a message was previously sent to an agent, but not received by it yet, this rule can be applied to simulate the reception of such a message.
The rule makes the agent a container of the message.
Hence, after applying this rule, the message acts as a guard, and its graphical representation changes.
\sect{Time Stepping Rule}
Contains a single rule \textit{Step Time}, not depicted due to its simplicity, which represents the forward passage of time for a given amount.

\sect{Remove Active Marks Rule}
Same as before, this group just contains a very simple rule, not shown, to set \textsf{active=false} in an agent, hence removing the mark.

\subsection{Coordination Algorithm}
The MCMT rules aforementioned must be coordinated with certain priorities in order to yield the expected result.
This process of coordination consists simply of a number of layers or priorities which we indicate with circled numbers (\circled{n}), to which each rule belongs.
Inside a group, we can either choose to run a single matching rule from that layer, or sequentially run rules from the same layer until no more rules match.
The coordination for our example scenario consists of the following layers:
\circled{1}~Run as many rules as possible from the \textit{Behavioural rules} until no further matching rules are found, thus allowing all agents to progress by reacting to environmental changes or incoming messages.
\circled{2}~Apply one rule from the \textit{Environmental Rules}.
If more than one rule matches the current snapshot, the user can manually select (hence guiding the execution), or the system will non-deterministically choose one of them.
\circled{3}~Run the \textit{Time Stepping} rule once.
This represents the global passage of time. 
\circled{4}~Apply the coordination (sub)algorithm for the rules for monitoring TLTL properties, which is given in the next section.
\circled{5}~Finally, apply the rule \textit{Remove Active Mark} as many times as possible, to remove it from all agents.
Once the last layer is reached, one new snapshot and its corresponding evaluation have been generated, and the process starts from the beginning again.
In Fig.~\ref{fig:example-execution}, the layers applied to generate a new snapshot are indicated with \textsf{B} (Behavioural), \textsf{E} (Environmental) and \textsf{T} (Time step).
The \textit{TLTL monitoring} and \textit{Remove Active Marks} layers are always executed after \textsf{T}.


	\vspace{-2mm}

\section{TLTL Modelling and Semantics}
\label{sec:ftltl-modelling-semantics}

The model representing the TLTL language consists of nodes representing the non-terminal symbols of the TLTL grammar, which are specialised using inheritance into the specific operators of the language.
The relations among operators in a property are expressed as edges, so that the subformula contained inside an operator can be a whole expression, recursively defined.
It is worth mentioning that all nodes inherit from the one called \textsf{Formula}, which contains a boolean attribute represented as \evaluateMark{} when set used during the property evaluation explained below.
A specific TLTL property is built as an instance of this model.
Due to the EBNF-like structure of this model, its instances are tree-shaped structures, with the outermost operator as the root of the tree, its subformulas as the children nodes of the root, the outermost operator of a subformula as the root of the subtree, and so on.
The property shown in Fig.~\ref{fig:example-evaluation} expresses that whenever the master encounters an obstacle, the slaves will stop within a given time interval (3 units) through the temporal pattern \(\LTLglobally (o \to (\LTLprophecy{0,3} m_1 \land \LTLprophecy{0,3} m_2)) \).

\subsection{Snapshots, Words and Evaluation}

\begin{figure}[h!]
	\centering
	\includestandalone[width=.7825\textwidth]{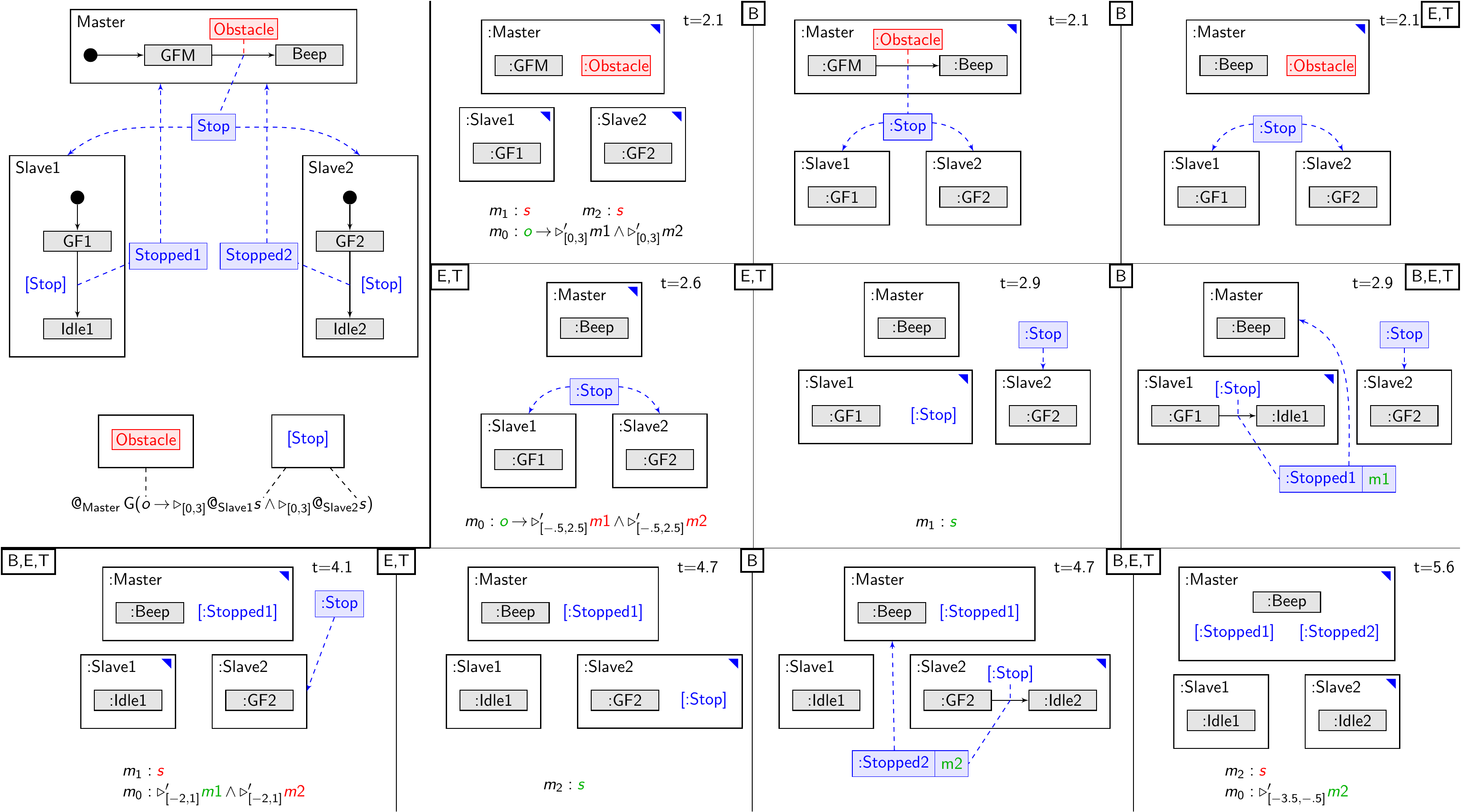}
	\caption{Example of simulation with the resulting snapshots}
	\label{fig:example-execution}
\end{figure}

\begin{figure}[h!]
	\centering
	\includegraphics[width=.925\textheight,angle=90,origin=c]{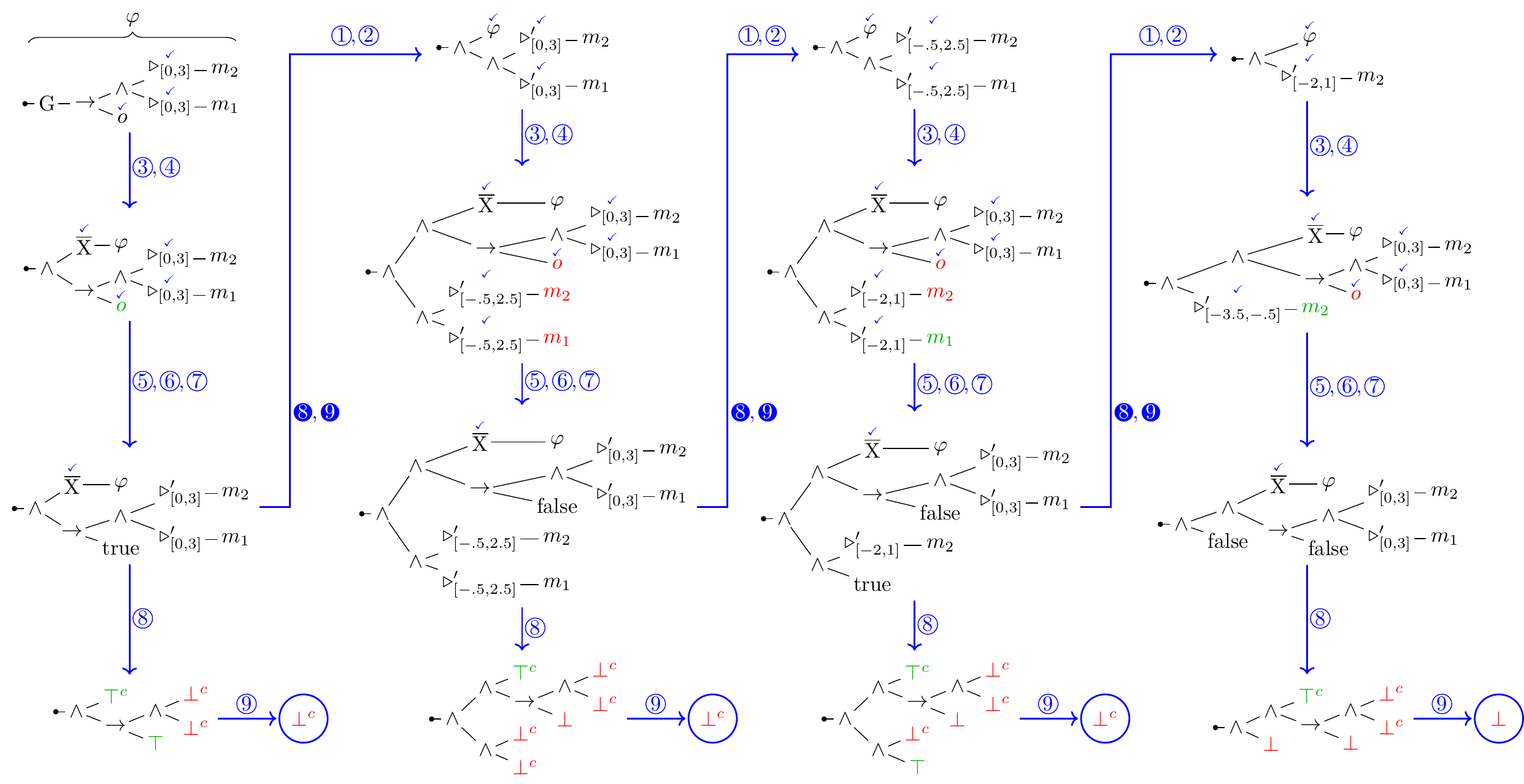}    
	\caption{Example of evaluation of a distributed, timed LTL property}
	\label{fig:example-evaluation}
\end{figure}

To check if an execution sequence of \textsf{Robolang} snapshots fulfils a TLTL specification, we need to relate a snapshot sequence to an input word for TLTL such that we can apply the TLTL semantics as defined before. In order to link a language such as \textsf{Robolang} with the TLTL formulas, the language must contain mappings for the following concepts:
Agents, to be able to relate monitors to the different actors in a distributed system;
an \emph{Active} mark in agents, to know when their associated monitor must (re)evaluate a formula;
Messages, for the distribution of evaluations in remote monitors (see the example in Figure~\ref{fig:example-execution} and~\ref{fig:example-evaluation});
and a clock, where each agent can retrieve relative time differences to calculate the time increase.
An \textsf{Agent} (white box) contains the \textsf{active} boolean mark (blue corner), as well as \textsf{Tasks} (grey boxes), \textsf{Transitions} (black arrows) that connect them and \textsf{Messages} (blue boxes) that can be send and received.
The \textsf{Messages} acting as guards (blue brackets), together with \textsf{Inputs} (red boxes) and \textsf{Timed Guards} (not used in our example) can trigger a \textsf{Transition} (dashed lines) to the next \textsf{Task}.

\pagebreak
Now we can interpret a snapshot sequence as word $w \in (\Sigma \times \mathbb T)^+$:
For a TLTL formula $@_A\varphi$ we have a word $w^A = (w^A_0,t^A_0)\dots (w^A_n,t^A_n)$ for every agent $A$ by having $w^A_i$ for the $i$-th snapshot that has an active mark in $A$ such that $w^A_i$ contains all propositions, whose linked model fragments are present in the snapshot, and $t^A_i$ is the current timestamp of the snapshot.

We apply online monitoring through evaluating the TLTL properties along with executing the \textsf{Robolang} models.
We implemented the TLTL monitoring function in terms of MCMT rules: 
We use unrolling for the currently to-be-evaluated temporal operators and then replace the propositions with the truth values according to the current snapshot and propositions, as well as adjusting the time of prophecy operators. 
Then, by replacing all currently evaluated operators by the corresponding truth values from the four-valued domain, we can calculate the current truth value resulting from the last seen snapshot. 
For the snapshots given in Fig.~\ref{fig:example-execution}, this leads to the trees in Fig.~\ref{fig:example-evaluation}. 

\subsection{Monitoring TLTL using MCMTs}
\setlength{\abovecaptionskip}{-2pt plus 0pt minus 0pt}
\begin{wrapfigure}[8]{r}{0.35\textwidth}
	\vspace{-5mm}
	\centering
	\includestandalone[scale=.53]{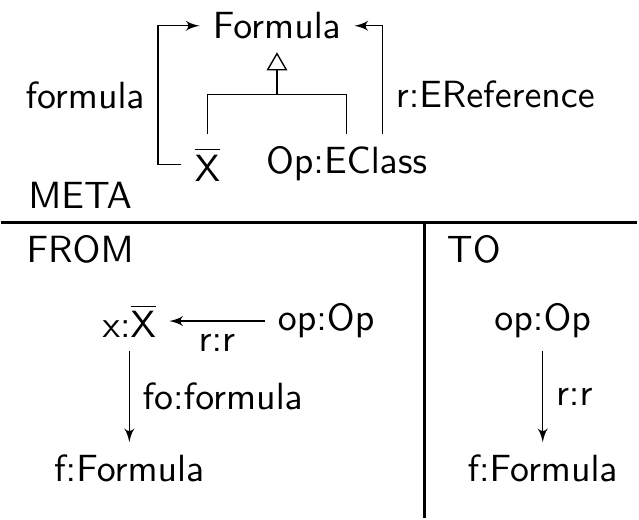}
	\caption{\small Delete Weak Next}
	\label{fig:ltl-mcmt-delete-weak-next}
\end{wrapfigure}
\setlength{\abovecaptionskip}{10pt plus 0pt minus 0pt}
Due to space limitations, we do not include an exhaustive list of all the MCMTs required to express the TLTL monitoring function defined in Section~\ref{sec:ltl}.
As an illustrative example, Fig.~\ref{fig:ltl-mcmt-delete-weak-next} depicts the \textit{Delete Weak Next} rule, used in step \bluecircled{8}.
The rest of them follow a similar structure, and are coordinated as follows (again representing layers as circles):
The first steps mark the outermost temporal operators:
\circledblue{1}~Set \evaluateMark{} in the root of the expression.
\circledblue{2}~As long as there is a boolean operator having \evaluateMark{}, remove \evaluateMark{} from it and set \evaluateMark{} to all of its children.

Now all the outermost temporal operators and atomic propositions are marked with \evaluateMark{}.
Next we apply unrolling operations for the marked temporal operators:
\circledblue{3}~For every temporal operator having \evaluateMark{}, the corresponding unrolling rule is applied according to the monitoring function described in Section~\ref{sec:ltl}.
The rules that generate operators or atomic propositions outside of temporal operators, mark these with \evaluateMark{}.
\circledblue{4}~Update all $\LTLprophecyobligation{t_1,t_2}$ with $\LTLprophecyobligation{t_1-\delta,t_2-\delta}$ where $\delta$ is the time difference between the current and the previous snapshot.
If a timed boundary becomes negative, it means that the boundary has been trespassed.

In the end, all the operators and atomic propositions outside of temporal operators are marked with \evaluateMark{}.
Next the prophecies and atomic propositions must be evaluated:
\circledblue{5}~Replace every atomic proposition having \evaluateMark{} with $\LTLtrue$ or $\LTLfalse$, depending on whether we find a match of the atomic proposition in the current snapshot.
\circledblue{6}~For every $\LTLprophecyobligation{t_1,t_2}$ having \evaluateMark{}, replace it with $\LTLtrue$ or $\LTLfalse$ according to the monitoring function described in Section~\ref{sec:ltl}.
If it remains unchanged, remove \evaluateMark{}.
\circledblue{7}~Replace every $\LTLprophecy{t_1,t_2}$ having \evaluateMark{} with $\LTLprophecyobligation{t_1,t_2}$.
These rules then remove \evaluateMark{} from the affected operators.

Next, the result of the previous steps is used for two different paths of transformations:
On one of them, the rules calculate the four-valued boolean evaluation of the property for the current state of the system, i.e. current snapshot:
\circledblue{8}~Replace $\LTLtrue$ by $\top$, $\LTLfalse$ by $\bot$, $\LTLnext\phi$ and $\LTLprophecy{t_1,t_2}\phi$ by $\bot^c$, and $\LTLweaknext\phi$ by $\top^c$.
All subformulas $\phi$ from these operators are removed, leaving the tree just with values from $\mathbb{B}_4$ and boolean operators.
\circledblue{9}~Apply operations on $\mathbb{B}_4$ to the resulting tree until the whole tree collapses to a single $\mathbb{B}_4$ value, which is the current evaluation of the property.

Independent from the previous two steps, the state of the property is also used to calculate the obligation property, i.e. what must be evaluated in the next snapshot.
These two steps are performed in a different branch of execution, and therefore represented as filled circles:
\bluecircled{8}~For every $\LTLnext\phi$ and $\LTLweaknext\phi$ having \evaluateMark{}, they get replaced with their subtree $\phi$.
\bluecircled{9}~Apply boolean simplifications to any feasible match found throughout the tree, e.g.\ $\LTLtrue \LTLand x \equiv x$.
The resulting model represents the obligation property to be evaluated in the next snapshot.

Fig.~\ref{fig:example-evaluation} displays the simplified model representation of the property evaluated by the monitor $m_0$ in the snapshots where the agent \textsf{:Master} is active, and its evaluation to ultimately \emph{false} ($\bot$).
The blue arrows between models represent the application of each group of rules defined above.

\section{Related Work \& Conclusion}
\label{sec:related}
We have focussed on a temporal logic which is well-known, intuitive (to some degree), and that does not distract with its complexity from the core idea of coupling simulation/execution and runtime verification.
The Linear Temporal Logic (LTL, \cite{ltl4}) and its extensions to timed and distributed properties \cite{dtl,tltl} makes it suitable for our task, but are of course not the only possible choice.
As we encode the verification step-wise via the transformation rules of our framework, other logics are equally suited.
We distinguish this clearly from approaches where verification is done via a separate tool that is connected through abstract events: while this may allow a wider choice of logics, we no longer have the advantage of having the property coupled to the actual model, but would rather require separate model transformation steps of translating the property from its term-representation into the input for the verification tool, as well as serialize events, and hence forgo the advantages of co-evolution between model and specification.

The verification community uses specialised tools such as Uppaal or CPNTools, but they constrain the modeller, as the focus is more on verification than on concepts and modelling.
A notable exception are rewriting logics-based approaches in Maude: models are encoded via sorts, which allows capturing a hierarchy of abstraction levels, and configurations that capture states.
Rewriting rules can then address orthogonal or overlapping concerns within a single configuration, tying together the ontological model with its behaviour rules, supplementary sorts with their own set of rules, and rules which map events in the model to input events to the logic \cite{DBLP:conf/kbse/HavelundR01}.

Incorporating evolution of the model over time into the modelling process is in itself not new.
Model transformation are already used for definition of behaviour~\cite{CsertanHMPPV02,LaraV02,taentzer2003agg,Rensink03,rivera2009graphical,Schurr2014}.
Model versioning and suitable techniques have been studied by Altmanninger et al.\cite{DBLP:journals/ijwis/AltmanningerSW09}.
Linear Temporal OCL (LT-OCL, \cite{DBLP:conf/ecmdafa/SodenE09}) is an approach to allow temporal queries over dynamic behaviour, for the purpose of defining reactive behaviour.
Gomez et al.\cite{10.1007/978-3-030-00847-5_26} have turned this idea into a working framework built on EMF, called TemporalEMF.
In this system, one would still need an additional layer of encoding e.g. for the binary temporal operators into the OCL-dialect.
Both in LT-OCL and TemporalEMF, one has to some degree the desired co-evolution through syntax-checking of the OCL-queries against the models.
Encoding a temporal logic in a temporal query language fixed by OCL, any EMF-based solution would still require additional logic and a flattened model to take into account the multiple levels of the specification, i.e.\ the temporal language can not be easily changed by swapping out a supplementary hierarchy.

\paragraph{\bf Conclusion and Future Work.}
In this paper we used the expressive power of MLM to independently specify a behaviour model and its rules for execution, and a logic and its semantics for verification.
Through double-typing, we couple atomic propositions in the logic with the state of the simulation.
Both the simulation rules and the evaluation rules for timed properties are given as MCMTs, and through proper coordination we perform runtime verification of a run of a model, i.e. evaluating a sequence of states wrt. the specified logical property.
The close link between the correctness properties and the models allows to keep both synchronized during refinement processes of the models and even for the deployment of the models, e.g. using code generation.
Since in RV we only verify one particular run, we have been able to integrate the progress of time and the behaviour of the environment as part of the execution.

We plan to extend the language of MCMTs with rule inheritance, which would avoid some repetition (e.g.\ the rules for firing transitions).
Another topic for future work would be how the usage of different real time logics would affect our approach and if more expressive logics like Metric Temporal Logic \cite{mtl} or Metric Interval Temporal Logic \cite{mitl} can be used.

{\tiny
\bibliographystyle{alphaurl}
\bibliography{bibliography}
}
\end{document}